\documentclass[12pt,preprint]{emulateapj}

\makeatother

\lefthead{Magdis et al..~}

\slugcomment{accepted in ApJ}
\shorttitle{Star formation Rates of LBGs at $z\sim$3. }
\shortauthors{Magdis et al.}
\begin{document}
\title{A Multi-wavelength View of the  Star Formation Activity at z$\sim$3}

\author{G.E. Magdis,$\!$\altaffilmark{1}
D. Elbaz,$\!$\altaffilmark{1}
E. Daddi,$\!$\altaffilmark{1}
G.E. Morrison,$\!$\altaffilmark{2,3}
M. Dickinson,$\!$\altaffilmark{4}
D.Rigopoulou,$\!$\altaffilmark{5}
R.Gobat $\!$\altaffilmark{1}
\& Ho Seong Hwang$\!$\altaffilmark{1}
}
\altaffiltext{1}{CEA, Laboratoire AIM, Irfu/SAp, F-91191 Gif-sur-Yvette, France}
\altaffiltext{2}{Institute for Astronomy, University of Hawaii, Honolulu, HI 968226}
\altaffiltext{3}{Canada-France-Hawaii Telescope, Kamuela, HI 96743}
\altaffiltext{4}{NOAO, 950 N. Cherry Avenue, Tucson, AZ 85719, USA}
\altaffiltext{5}{Department of Astrophysics, Oxford University, Keble Road, Oxford, OX1 3RH}

\begin{abstract}
We present a multi-wavelength, UV-to-radio analysis for a sample of massive (M$_{\ast}$ $\sim$ 10$^{10}$ M$_\odot$) IRAC- and MIPS 24$\mu$m-detected 
Lyman Break Galaxies (LBGs) with spectroscopic redshifts z$\sim$3 in the GOODS-North field (L$_{\rm UV}$$>$1.8$\times$L$^{\ast}_{z=3}$). For LBGs without individual 24$\mu$m detections, we employ stacking techniques at 24$\mu$m, 
1.1mm and 1.4GHz, to construct the average UV-to-radio spectral energy distribution and find it 
to be consistent with that of a Luminous Infrared Galaxy (LIRG) with L$\rm_{IR}$=4.5$^{+1.1}_{-2.3}$$\times$10$^{11}$ L$_{\odot}$ 
and a specific star formation rate (SSFR) of 4.3 Gyr$^{-1}$ that corresponds to a mass doubling time $\sim$230 Myrs. 
On the other hand, when considering the 24$\mu$m-detected LBGs we find among them galaxies with
L$\rm_{IR}>$10$^{12}$ L$_{\odot}$, indicating that the space density of $z\sim$3 UV-selected 
Ultra-luminous Infrared Galaxies (ULIRGs) is $\sim$(1.5$\pm$0.5)$\times$10$^{-5}$ Mpc$^{-3}$. 
We compare measurements of star formation rates (SFRs) from data at different wavelengths
and find that there is tight correlation (Kendall's $\tau >$ 99.7\%) and excellent agreement between the values derived from dust-corrected UV, 
mid-IR, mm and radio data for the whole range of L$\rm_{IR}$ up to L$\rm_{IR}$ $\sim$ 10$^{13}$ L$_{\odot}$. This range is greater than that for which the correlation is known to hold at z$\sim$2, possibly due to the lack of significant contribution from PAHs to the 24$\mu$m flux at $z\sim$3. The fact that this agreement is observed for galaxies with L$\rm_{IR}$ $>$ 10$^{12}$ L$_{\odot}$ suggests that star-formation in UV-selected ULIRGs, as well as the bulk of star-formation activity at this redshift, is not embedded in optically thick regions as seen in local ULIRGs and submillimeter-selected galaxies at $z=2$.

\end{abstract}

\keywords{cosmology: observations --- galaxies: evolution --- galaxies: high redshift ---infrared: galaxies}

\begin{figure*}[!t]
\centering
\includegraphics[width=15cm,height=10cm,angle=0]{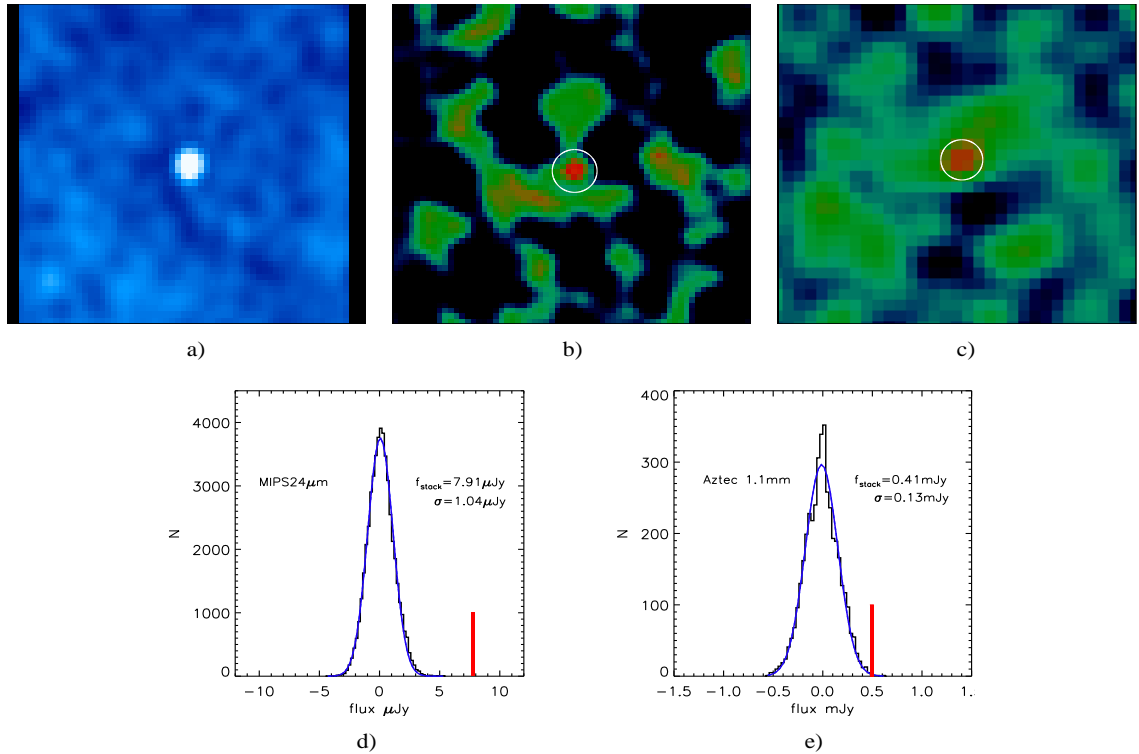}
\caption{Stacked image at 24$\mu$m (a), 1.1mm (b) and radio 1.4GHz (c) of IRAC-LBGs d) Stacking simulations at 24$\mu$m. Distribution of the measured fluxes derived from 50.000 stackings at 40 random positions along with the best gaussian fit (rms=1.04$\mu$Jy). The red line denotes the flux measurement of the stacking at the position IRAC-LBGs. This indicates a $\sim$7.9$\sigma$ detection at the stacked position of the LBGs. e) Stacking at 1.1mm. Same as in d) but for the Aztec 1.1mm and for 5000 stackings. This figure indicates a $\sim$3.7$\sigma$ detection at the stacked position of the IRAC-LBGs.}
\label{fig:sub} 
\end{figure*}

\section{Introduction}
One of the most fundamental quantities needed for understanding the nature and evolution of galaxies is the star formation rate (SFR). To get reliable and meaningful estimates of the SFR for galaxies at high redshift, one needs a well defined sample of objects, coupled with 
multi-wavelength data that can provide a thorough and  comprehensive investigation. 

One of the most successful methods of detecting high--z star--forming galaxies is the Lyman-break technique, pioneered by Steidel et al.\ (1996,2003). This technique has revealed a wealth of Lyman Break Galaxies (LBGs) at $z\sim$3, now comprising an impressive catalogue of thousands star--forming galaxies at this redshift. 
Multi-wavelength studies of LBGs have provided extensive information on various
physical properties of these objects.   In particular, measurements at near-infrared wavelengths
and at 3.6-8$\mu$m from the Spitzer Space Telescope IRAC instrument indicate that
their stellar masses are typically 10$^{9}$-10$^{11}$ M$_{\odot}$ (e.g., Shapley et al.\ 2001, Papovich et al.\ 2001, Magdis et al.\ 2010). 

The dust content and the SFR of LBGs at $z \approx 3$ are still poorly constrained. 
For their siblings at lower redshift, $z\sim$2, Reddy \& Steidel (2004) and Reddy et al.\ (2006), using 
multi-wavelength data ranging from X-rays to radio, have reported that UV can be a reliable SFR indicator if corrected for dust attenuation by an average factor between 4.4 and 5.1. The validity of the UV as a robust SFR indicator has also been presented by Daddi et al.\ (2005, 2007) for a sample of near-infrared selected galaxies at 1.5 $<$ $z$ $<$ 2.5 identified using the BzK technique (see also Dannerbauer et al.\ 2006). A similar multi-wavelength study for the $z\sim$3 LBGs, though, is still needed.

In this letter we make use of the unique compilation of multi-wavelength data on the Great Observatories Origins Deep Survey North field (GOODS-N)
to explore the SFR and the infrared luminosities (L$\rm_{IR}$) of $z\sim$3 LBGs. Our aim is to fully characterize
the spectral energy distribution (SED) of a typical LBG from rest-frame UV to radio wavelengths, to compare different tracers of star formation, and to test whether the UV can provide a reliable measurement of star formation at $z\sim$3. For this letter we adopt a $\Lambda$ cold dark matter ($\Lambda$CDM) cosmology with  H$_{0}$= 71 km s$^{-1}$ Mpc$^{-1}$, $\Omega_{m}$= 0.27 and  $\Omega_{\Lambda}$= 0.73, while the magnitudes presented in this work are all in  the AB magnitude system.

\section{The LBG sample and Data Sets}
\subsection{Rest-frame UV data}
In this study we selected 69 LBGs in the northern field of the Great Observatories Origins Deep Survey (GOODS). Their original selection was based on their optical colours ($U_{n}$,$G$,$R$) to $R=$25.5 by Steidel et al.\ (2003) and subsequent optical spectroscopy (Steidel et al.\ 2003, Reddy et al. 2005) has confirmed their high redshift nature with a median $z$=2.95. Optical spectroscopy has also been used to determine the absence of AGN signatures (i.e., strong high ionization emission lines) in their rest-frame UV spectrum, although a deeply obscured AGN cannot be ruled out from these data. We also use $BViz$ data obtained from the Advanced Camera for Surveys (ACS) on board the Hubble Space Telescope (HST) (Giavalisco et al.\ 2004). 

\subsection{IRAC and MIPS 24$\mu$m data}
Rest-frame NIR identification and IRAC photometry (5$\sigma$, $f_{3.6}$=0.21 $\mu$Jy) of the LBGs in our sample have been presented by Magdis et al.\ (2008,2010), who also showed that their mid-IR colours are consistent with those of star-forming galaxies at $z\sim$3. Here, we focus on the 49  LBGs with at least one IRAC detection ([3.6]$\rm_{AB}<$25.0), in order to place a lower mass limit of M$_{\ast}\sim$10$^{10}$M$_\odot$ (Magdis et al. 2010) and facilitate a robust investigation of their properties and their average SED. We not the R magnitude limit of our sample, ($R_{\rm AB}<$25.5), corresponds to 
L$_{\rm UV}$ $>$ 1.8$\times$L$^{\ast}_{z=3}$  (M$^{\ast}_{\rm R}$=-21, Steidel et al. 1999)

We matched our sample with a GOODS-N 24$\mu$m catalog (5$\sigma\label{•} \sim 20 \mu$Jy, translated to SFR$\approx$350 M$_{\odot}$yr$^{-1}$ or L$\rm_{IR}\approx$2$\times$10$^{12}$ L$_\odot$ based on Chary \& Elbaz 2001 models at z=3) produced by the GOODS team (Dickinson et al.\ in prep.), and searched for counterparts within a 2'' diameter separation
centered on the optical position. We identify 9 LBGs and we add another five 24$\mu$m-detected LBGs in the Extended Growth Strip (EGS, 5$\sigma\label{•} \sim$70 $\mu$Jy) by Rigopoulou et al.\ (2006), to increase our 24$\mu$m-detected sample. We note that the original optical selection of these extra 
five EGS MIPS-selected LBGs is identical to that of the GOODS-N sample, so they share similar UV properties and including them in our sample doesn't introduce any bias. Henceforth, we will refer to LBGs that are individually detected at 24$\mu$m as MIPS-LBGs (14 objects), and those that are not as IRAC-LBGs (40 objects). In practice, all MIPS-LBGs are also detected in all four IRAC bands. The redshift range of the two sample is 
2.63$<z<3.41$ (IRAC-LBGs) and 2.60$<z<3.31$ (MIPS-LBGs) following a similar distribution. The median 
redshift of the MIPS- and IRAC-LBGs is 2.92 and 2.98 respectively.

For the 24$\mu$m-undetected IRAC-LBGs in GOODS-N we employed median stacking analysis. We first subtracted all detected sources using the PSF 
used for the source extraction, and cut sub-images centered at the optical position of each undetected LBG. 
To avoid contaminating the stacked signal from residuals, we only added galaxies to the stack if there were no 
bright MIPS sources within $\sim$4'' of those galaxies. Then a stacked flux was measured in a manner 
similar to the measurement of the detected MIPS sources. The final stacked image at the position of the IRAC-LBGs is shown in Figure 1a. To quantify the error of our measurement we 
stacked at 40 random positions and repeated it 50.000 times. As expected for white noise, the distribution of the fluxes follows a gaussian shape with an rms of 1.04 $\mu$Jy (Figure 1d) that we adopt as the uncertainty of our measurement. 
 The median flux density of the IRAC-LBGs as derived from stacking is $f_{24}$=7.91$\pm$1.04$ \mu$Jy (S/N $\sim$8). 

Previous studies have demonstrated that MIPS-LBGs are, on average, more massive, relatively older and dustier compared to LBGs that are undetected at 24$\mu$m (e.g., Rigopoulou et al.\ 2006, Magdis et al.\ 2010). In the rest of this paper we study the IRAC-LBGs and MIPS-LBGs separately to enable a comparison between the two sub-populations of LBGs and examine our results as a function of L$\rm_{IR}$ and dust extinction. For our statistical analysis we always refer to median values.

\subsection{Aztec 1.1mm data}
Recently, a deep ($\sigma$ $\sim$ 0.96-1.16 mJy beam$^{-1}$) and uniform 1.1mm survey of the GOODS-N field with AzTEC (Wilson et al.\ 2008) was conducted on the James Clerk Maxwell Telescope. Matching our sample with the catalogs published by Perera et al.\ (2008) and Chapin et al.\ (2009), returned no individual detection down to 3.75$\sigma$ level. Since none of our LBGs is individually detected we used the publicly available maps to do stacking. We perform median stacking at the positions of IRAC-LBGs after rejecting two of them that are separated by less than an Aztec beam half-width (9'') from any known Aztec detection (Figure 1b). The median flux that we recover for the IRAC-LBGs is $f\rm_{1.1mm}$=0.41$\pm$0.11 mJy (3.7$\sigma$). The robustness of this flux was also tested by the same method employed for the 24$\mu$m stacking (Figure 1e) and bootstrapping. Finally, stacking at the positions of the seven GOODS-N MIPS-LBGs (two were rejected due confusion) returned no detection indicating a 4$\sigma$  upper flux density limit of 1.08 mJy.   

\subsection{VLA 1.4GHz  radio data}
We use a new deep 20 cm (1.4 GHz) imaging
of the GOODS-N field obtained at the VLA in ABCD configurations
(G. Morrison et al.\ 2010, submitted). Previous observations by
Richards (2000) contained 40 usable hours of A-array data which was
supplemented with 125-hr yielding an rms of 3.9$\mu$Jy beam$^{-1}$ near
the phase center. We matched our LBG sample to the 5$\sigma$ radio
catalog (Morrison et al.\ 2010) and identified one MIPS-LBG, HDFN-M23 with a
flux S$_{\rm1.4GHz}$=21.19$\pm$4.2 $\mu$Jy ($z$=3.21).

In order to reach deeper radio flux densities, we stacked individually
undetected LBG sources using the following technique. Using a primary
beam corrected radio image, we made 100pixel $\times$ 100pixel sub-images
centered on the LBG positions. The exported FITS files were then
stacked using the Terapix co-add software Swarp (Bertin et
al.\ 2002). The stacking used the pixel reference frame and the
resulting stacked image was median combined. We stacked separately MIPS- and IRAC-LBGs and derived a median flux density of 8.5$\pm$2.2 $\mu$Jy and 3.6$\pm$0.8 $\mu$Jy respectively. The stacked image of the IRAC-LBGs is shown in Figure 1c.

\section{SFR indicators}
\textbf{UV SFR estimates.} To determine the UV-corrected SFR (SFR$\rm_{UV cor}$), 
we use the GOODS ACS photometric catalog retrieved from MAST STScI (Version v2.0). At $z\sim$3 the observed V and I and z bands correspond to
rest--frame 1500 $\rm\AA$, 2000 $\rm\AA$ and 2400 $\rm\AA$ respectively, allowing a robust estimate of the $\beta$ slope. We use dust free models of continuous SFR, solar metallicity, age t$\rm_{sfr}=$ 100Myrs and Salpeter IMF, generated with the new code of Charlot \& Bruzual 2007 (CB07 private com.), to fit the SED of each individual LBG after correcting for reddening using the Calzetti (2000) attenuation law and correcting for the IGM attenuation using the prescription of Madau (1995). From the best fit model, we derive E(B-V) values and apply a K-correction to infer the observed and subsequently the intrinsic flux density at 1500$\rm\AA$ in the rest frame. The average E(B-V) value is 0.16  (A$\rm_{V}$=4.6) and agrees well with previous studies (i.e., Shapley et al.\ 2001). The adopted CB07 models yield a relation between the SFR and the monochromatic 1500$\rm\AA$ luminosity given by:

\begin{equation}{
{\rm SFR (M_\odot{\rm yr}^{-1}) =  L_{1500 \AA}  [{\rm erg\ s^{-1} Hz^{-1}}] / (8.85\times10^{27})}}
\end{equation}
Using extinction-corrected L$\rm_{1500 {\mathrm \AA}}$ in equation 1, provides an estimate of the total SFR (SFR$\rm_{UV,corrected}$). The total SFR can be considered as the sum 
of an unobscured component (SFR$\rm_{UV,uncorrected}$), which can be computed from equation 1
by using the observed  L$\rm_{1500 \AA}$, and an obscured component
(SFR$\rm_{obsc}$), corresponding to the energy absorbed by dust.  This absorbed energy is in turn
reradiated at mid- and far-infrared wavelengths, SFR$\rm_{obsc}$ = SFR$\rm_{IR}$, and it is this latter quantity that we will infer from the mid- and far-infrared data. Therefore, we may write:
\begin{equation}{
{\rm SFR}_{\rm UV,corrected} = {\rm SFR}_{\rm IR} + {\rm SFR}_{\rm UV,uncorrected}}
\end{equation}
\textbf{24$\mu$m, 1.1mm, 1.4GHz radio SFR and L$\rm_{IR}$ estimates.} To convert 24$\mu$m and Aztec fluxes to total (8-1000$\mu$m) L$\rm_{IR}$, we use the luminosity-dependent SED library
of Chary \& Elbaz (2001) (CE01), while for comparison we also consider the Dale \& Helou 2002 (DH02) models (with luminosities normalized as described by Marcillac et al. 2006) and SED templates of Arp220 and M82. This is done by interpolating
 L$\rm_{IR}$ over the template SEDs, sorting the value of L$\rm_{IR}$ that corresponds to the observed 24 $\mu$m flux density. For the radio-based L$\rm_{IR}$ estimates, we first derive the observed radio luminosities and then assume a radio spectral index of $\alpha$=-0.8 to get the rest-frame 1.4GHz radio luminosities. Then we use the local radio--IR correlation (Condon 1992) to determine L$\rm_{IR}$ :
\begin{equation}
\centering
L_{\rm IR}/L_{\odot} = 3.5\times 10^{-12} L(\rm 1.4~GHz) \quad \rm [W Hz^{-1}]
\label{eq:radioIR}
\end{equation}

The L$\rm_{IR}$ is subsequently converted to SFR$\rm_{IR}$ using Kennicutt (1998) :
\begin{equation}{
{\rm SFR}_{\rm radio} \rm {[M_\odot yr^{-1}]} = 1.73\times10^{-10} L_{\rm IR} \rm [L_\odot].}
\label{eq:LIR}
\end{equation} 

\begin{figure}[!h]
\centering
\includegraphics[width=8cm]{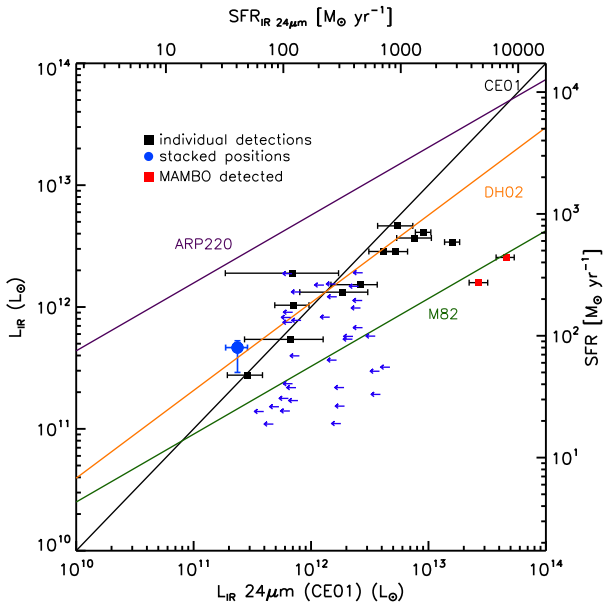}
 \caption{A comparison between the 24$\mu$m inferred luminosities and those derived from the obscured UV-light. The blue circle shows the stacking result for IRAC-LBGs, black squares show LBGs with 24$\mu$m detection (MIPS-LBGs) while red squares show MIPS-LBGs from EGS which have a 1.2mm  MAMBO detection. Blue arrows indicate 24$\mu$m upper limits (3$\sigma$) for the IRAC-LBGs. For these symbols, y-axis indicates L$\rm_{IR}$ (left) and SFR, both derived from UV uncorrected measurements. The black line shows the one-to-one relation based on CE01. The colored lines show the correlation between CE01 and DH02, ARP220 and M82. For these lines the left vertical axis should be read as the L$\rm_{IR}$ predicted from the corresponding models.}
\end{figure}

\section{Comparison of UV, mid-IR, mm and radio SFR/L$\rm_{IR}$ estimates}
\subsection{UV vs 24$\mu$m} 
We first compare the SFR and L$\rm_{IR}$ derived from UV and mid-IR in Figure 2, where we also show the expected correlation based on the CE01, DH02, Arp220 and M82 templates. For the IRAC-LBGs,  we use the L$\rm_{IR}$ derived from the stacking analysis, while for the rest we use the one derived from their measured 24$\mu$m flux densities. It is evident that between UV- and 24$\mu$m derived luminosities based on CE01 and DH02 templates, there is a close agreement which for the case of the latter templates is more prominent. The Kendall's $\tau$ test detects a correlation at a $>$99.7\% confidence level, that holds up to  L$\rm_{IR}\sim$10$^{13}$ L$_\odot$, where the two estimates appear to deviate. The two galaxies with the largest 24$\mu$m-derived L$\rm_{IR}$ and with the largest deviations between the UV and 24$\mu$m-derived SFR, are also detected at 1.2mm by MAMBO (Rigopoulou et al., in prep.), and are therefore members of the SMG population for which such a trend is known to exist either due to star formation embedded in optically thick regions or due to contribution from an AGN to the mid-IR output, or both. (e.g., Chapman et al.\ 2005, 2009, Pope et al. 2008).  We also note that for these two galaxies, a scaled-up M82 template provides a good agreement between the UV- and 24$\mu$m derived L$\rm_{IR}$.

A similar correlation between UV and mid-IR derived SFRs has also been observed at $z\sim$2 for BzK (Daddi et al.\ 2007)  and BX/BM (Reddy et al.\ 2006) galaxies with the two estimates though deviating at lower luminosities ($\sim$3$\times$10$^{12}$ L$_{\odot}$) when compared to our findings at z$\sim$3. It is therefore indicated that the UV-24$\mu$m correlation at z$\sim$3 is better than that found at z$\sim$2. One possible explanation could reside in the contribution of PAH features in the observed mid-IR flux. While evidence was found that PAHs are enhanced in z$\sim$2 star forming galaxies (Murphy et al. 2009) with respect to local galaxies, this may not be the case anymore at z$\sim$3 when galaxies were less metal rich and PAHs less abundant. We note that while for z$\sim$2 galaxies the 24$\mu$m is centered at the PAH emission, at $\sim$3 the band traces only the 6.2$\mu$m and part of the 7.7$\mu$m PAH feature. This shift though, cannot fully explain our findings as for a M82-like object, the contribution of the PAHs to the the 24$\mu$m flux density is comparable at the two redshifts (50-70$\%$). Alternatively, it could be argued that the mid-IR radiation of our sample is less polluted by the emission of hot dust heated by an AGN, which could be a natural result of the rapid drop of the number density of AGNs above z=2 (Wall et al. 2005).

\begin{figure}[!b]
\centering
\includegraphics[width=70mm,angle=0]{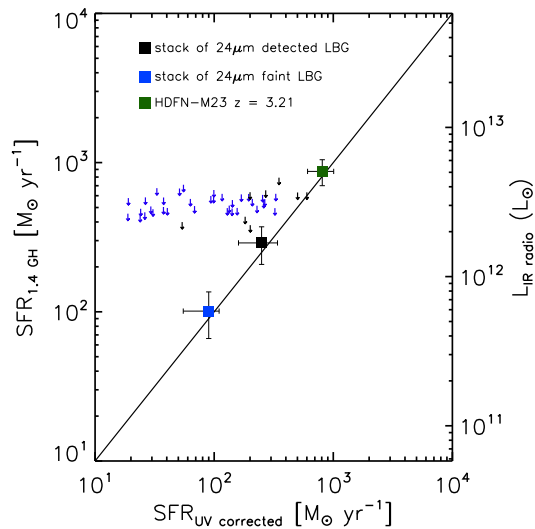}
\caption{A comparison between the radio 1.4GHz inferred SFR and those derived from the UV light from star--formation corrected for obscuration. Blue square shows the radio stacking of the IRAC-LBGs, black square shows radio stacking of MIPS-LBGs  while green square shows estimates for HDFN-M23, an individually detected LBG in the radio map. Straight line shows one to one correlation. The arrows indicate upper limits for the IRAC- (blue) and MIPS-detected LBGs.}
\label{fig:sub} %
\end{figure}

\subsection{UV vs radio and 1.1mm}
One independent way to check the UV SFR estimates is to compare them against radio observations. In Figure 3, we plot the SFR estimates based on the radio fluxes and the UV corrected for dust extinction for three samples: HDFN-M23 which is individually detected in the radio map, the stacked flux of radio-undetected MIPS-LBGs, and the stacked flux of radio-undetected IRAC-LBGs. We see that there is an excellent agreement between the radio and UV estimates, testifying to the validity of UV as a SFR indicator for UV-selected galaxies at $z\sim$3. 
The corresponding SFRs derived from UV (and radio) are 90$^{+20}_{-40}$ M$_{\odot}$yr$^{-1}$,(96$\pm$32 M$_{\odot}$yr$^{-1}$) 
for the IRAC-LBGs, 250$^{+35}_{-80}$ M$_\odot$yr$^{-1}$,(280$\pm$85 M$_\odot$yr$^{-1}$) for the MIPS-LBGS and 
808 M$_{\odot}$yr$^{-1}$,(870$\pm$200 M$_{\odot}$yr$^{-1}$) for HDFN-M23. A similar study by Carilli et al.\ (2008), comparing radio and UV data for $z \sim 3$ LBGs in COSMOS, inferred
an average UV attenuation factor of $\sim 1.8$, smaller than the average value of $\sim 5$ derived here and in other studies (e.g., Reddy \& Steidel 2004; Reddy et al.\ 2006).
We argue that the discrepancy arises from the fact that we focus on massive (dusty) LBGs with robust spectroscopic redshifts and compute dust attenuation on an object-by-object basis from the UV spectral slopes, whereas Carilli et al.\ primarily use photometric redshifts, infer only an average attenuation and do not consider a mass-limited sample.
\begin{figure*}[!t]
\centering
\includegraphics[width=15cm]{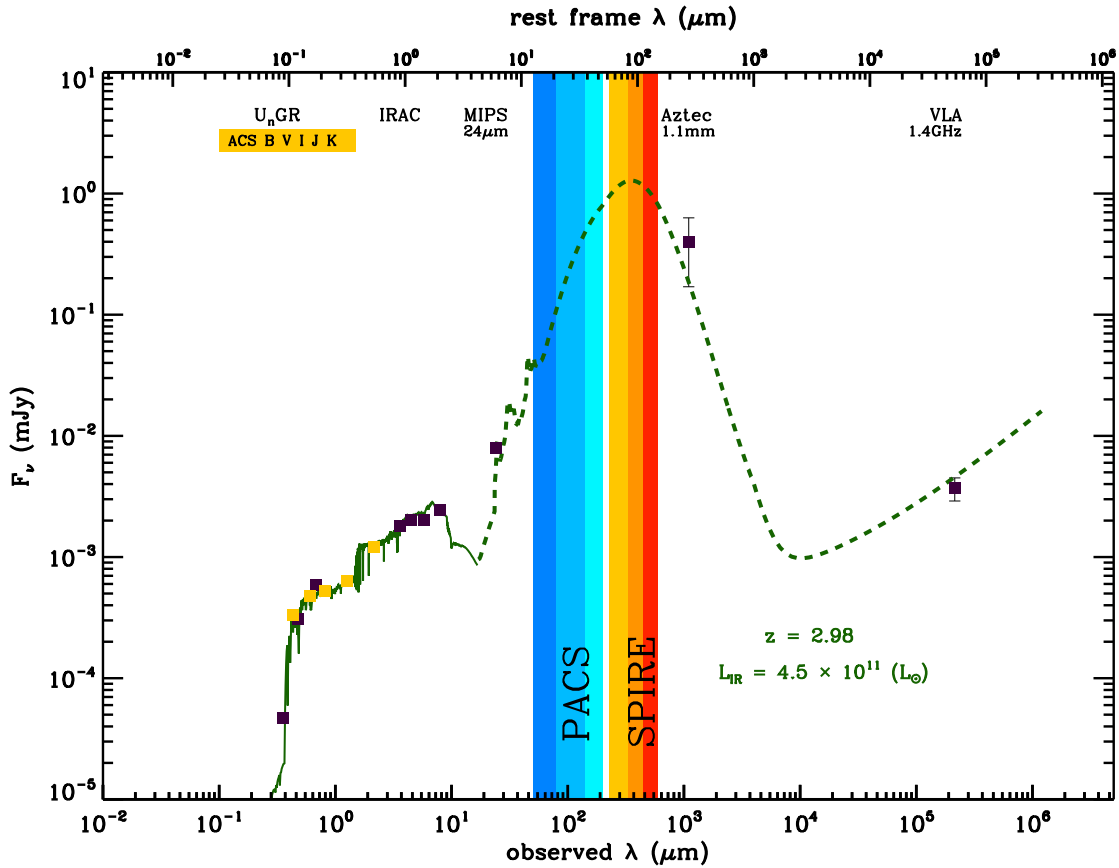}
 \caption{Average SED of a typical IRAC-detected LBG at $z\sim$3. For the SED we use the median BViJK+IRAC photometry of IRAC-LBGs, and the values derived from stacking MIPS24, Aztec and radio. The rest-frame UV-NIR portion of the data is overlaid with the best fit CB07 model, while the mid-IR to radio is shown with the best--fit CE01 model. The colored bands indicate the wavelength range sampled by the PACS and SPIRE instrument onboard the Herschel Space Observatory. }
\end{figure*}
Finally, we explore the relation between the  UV and the L$\rm_{IR}$ as derived from mm. Following the same prescription as above, we convert the mm stacked flux the IRAC-LBGs and find it to be in agreement with the L$\rm_{IR}$ derived from the UV. The two corresponding values are (6.2$\pm$3$)\times$10$^{11}$ L$_\odot$ (mm) and 5.1$\times$10$^{11}$ L$_\odot$ (UV). Converting the stacked mm value to SCUBA850$\mu$m  by using the formula presented by Ivison et al.\ (2005) we get $S_{850}$=0.85$\pm$0.27mJy indicating that IRAC-LBGs emit at the sub-mJy level at 850$\mu$m. We note that similarly to the SMGs, where UV is an unreliable indicator of SFR,  there is evidence of existing optically thick star-forming regions in the case of sub-mm detected LBGs (e.g., Rigopoulou et al.\ 2010 in prep, Chapman \& Casie 2009).

\section{Discussion}
Considering the median fluxes of the IRAC-LBGs for the rest--frame UV to NIR  (i.e U$_{n}$ to 8.0$\mu$m) and stacked fluxes at 24$\mu$m, 1.1mm and 1.4GHz, we construct the average SED of a typical (24$\mu$m faint) IRAC-LBG. We fit the rest--frame UV to NIR with model SEDs generated using the CB07 code, and the mid-IR to radio with CE01 templates. The photometric points along with the best fit model are shown in Figure 4. The best-fit CB07 model indicates an average stellar mass of M$_{\ast}$ $\sim$2.2$\times$10$^{10}$M$_{\odot}$ and an average SFR $\sim$85 M$_{\odot}$yr$^{-1}$. Using this mass estimate and the SFR derived from our multi-wavelength analysis we derive a specific SFR (SSFR, defined as SFR/M$_{\odot}$) $\sim$4.3 Gyr$^{-1}$, corresponding to a mass doubling time of  $\sim$ 230Myrs. This value is very close to the one presented by Magdis et al.\ (2010) (4.5 Gyr$^{-1}$) and is larger than that found at lower and higher redshifts, reinforcing their argument that the evolution of the SSFR peaks at $z\sim$3. Based on CE01 models we find that the average IRAC-LBG is a LIRG with L$\rm_{IR}$=4.5$^{+1.1}_{-2.3}\times$10$^{11}$ L$_{\odot}$. We note that L$_{\rm IR}$ corresponds to a typical dust temperature T$_{\rm d}$=35K, based on T$_{\rm d}$ measurements of local LIRGs (Yang et al. 2007).

On the other hand, MIPS-LBGs have higher luminosities, indicating that ULIRGs are present among the UV-selected galaxies at $z\sim$3. For instance, HDFN-M23 which was individually detected in the radio, has L$\rm_{IR}\sim$5($\pm$2)$\times$10$^{12}$ L$_{\odot}$. We calculate the comoving volume for the redshift range of our sample (2.5$<$ z $<$3.5) and find that the space density of ULIRGs (based on UV L$\rm_{IR}$) LBGs is  $\sim$1.5$\pm$0.5$\times$10$^{-5}$ Mpc$^{-3}$. This is a factor of $\sim$10 smaller than the space density of the $z\sim$2 BzK-selected ULIRGs (Daddi et al.\ 2007). We note that the UV selection is likely to miss more obscured, UV faint ULIRGs as our sample is limited to R$_{\rm AB}$ $<$ 25.5 while, for example, an Arp220 like object at $z\sim$3 would have R$_ {\rm AB}\sim$26.9.

The good agreement between SFR indicators that are affected by dust extinction (UV) or not (MIR, mm, radio) suggests that the bulk of the star formation activity in massive UV-selected galaxies takes place in optically thin regions. Since by focusing on the massive LBGs we also select those that are most affected by dust extinction (e.g., Magdis et al.\ 2010), it is reasonable to extend this result to the whole population of UV galaxies including the less massive, less dusty LBGs. Furthermore, the fact that the agreement holds even for the case of the most massive, dusty, 24$\mu$m-detected LBGs with  L$\rm_{IR}>$10$^{12}$ L$_{\odot}$, indicates that UV-selected ULIRGs at z=3 are transparent to UV light, meaning that we can estimate its SFR given its rest-frame UV 1500 $\rm \AA$ luminosity and UV slope, contrary to the local ULIRGs and z=2 SMGs. Similar results have been reached by Reddy et al. (2006) for UV-selected and by Daddi et al. (2007) for BzK selected  z$\sim$2 LIRGs and ULIRGs. Both studies find that galaxies of a given bolometric luminosity are on average a 
factor of 8-10 less dust obscured at z$\sim$2 than at the present epoch. The fact that our study suggests that at z$\sim$3 ULIRGs are optically thin at even higher L$_{\rm IR}$ than that at z$\sim$2, indicates even less obscuration at z$\sim$3 confirming the trend  between galaxies at z = 0, z $\sim$ 1, and z $\sim$ 3 (Adelberger $\&$ Steidel 2000). This could plausibly be explained as a 
result of increasing dust-to-gas ratios as we move from the high-z to the present universe. As galaxies evolve, they convert gas into stars which in turn enrich the interstellar medium with dust. If the dust distribution becomes more compact with time (assuming mergers that drive dust and gas to the center of the galaxy) the overall result would be an increase of the dust column density toward star-forming regions making ULIRGs  progressively optically opaque at later epochs.

To put constraints on the census of the cosmic star formation at z=3, we should also consider the missing fraction of star formation embedded in optically thick regions, that takes place in galaxies not selected in UV. On the other hand, there is evidence of decreasing obscuration with increasing redshift for a given L$\rm_{IR}$ (e.g., Reddy et  2008), pointing towards less optically thick star-formation at higher redshifts. Combining the above with the fact that the bulk of the currently known SMGs are at $z\sim$2.2 (Chapman et al.\ 2005), we can assume that at z=3 the contribution of the SMGs to the total SFR density, is not dominant, and hence locate the census of high-z star-formation in optically thin regions. We stress, that such a scenario cannot be confirmed based solely on our sample, as it has been shown that the Lyman Break technique can miss a large fraction of massive (dusty) galaxies at high-z (e.g., Daddi et al. 2004, Van Dokkum et al.\ 2006). 

Further insights into the far-IR properties of the LBGs, such as dust temperature and dust mass will be provided by deep surveys with the Photodetector Array Camera \& Spectrometer (PACS, 70-,100-,160$\mu$m) and the Photometric Imaging REceiver (SPIRE, 250-,350-500$\mu$m) on board the Herschel Space Observatory (HSO). Based on the average spectrum of  Figure 4, the predicted flux densities in the HSO bands of a typical IRAC-detected LBG are $f_{100}$=0.13 mJy, $f_{160}$=0.47 mJy, $f_{250}$=1.18 mJy, $f_{350}$=1.52 mJy and $f_{500}$=1.22 mJy.

%\begin{figure*}
%\centering
%\includegraphics[width=65mm,angle=0]{uv-radio.eps}
%\vspace{0.2cm}
%\hspace{0.2cm}
%\includegraphics[width=57mm,angle=0]{tt.eps}
%\caption{\textbf{left)} A comparison between SFRs inferred from 1.4GHz radio data and those derived from the UV light corrected for obscuration. The blue square shows the radio stacking of the 24$\mu$m-undetected LBGs, the black square shows radio stacking of LBGs with 24$\mu$m detection, and the green square shows the measurement for HDFN-M23 which is individually detected in the radio map. Straight line shows one to one correlation. The arrows indicate upper limits for the 24$\mu$m faint (blue) and 24$\mu$m-detected (black) LBGs. \textbf{right)} A comparison between the Aztec 1.1$\mu$m inferred luminosities and those derived from the obscured UV light from star--formation. The blue square shows  the estimates based on the stacking of 24$\mu$m faint LBG while black and red squares show the upper limits for aLBGs and 24$\mu$m LBGs with L$\rm_{IR}$$>$10$^{12}$L$_{\odot}$ respectivelly.}
%\label{fig:sub} %
%\end{figure*}

\end{document}